\documentclass[10pt,twocolumn,letterpaper]{article}

\usepackage{iccv}
\usepackage{times}
\usepackage{graphicx}%
\usepackage{multirow}%
\usepackage{amsmath,amssymb,amsfonts}%
\usepackage{amsthm}%
\usepackage{mathrsfs}%
\usepackage[title]{appendix}%
\usepackage{xcolor}%
\usepackage{textcomp}%
\usepackage{manyfoot}%
\usepackage{booktabs}%
\usepackage{algorithm}%
\usepackage{algorithmicx}%
\usepackage{algpseudocode}%
\usepackage{listings}%
\usepackage[utf8]{inputenc}
\usepackage[T1]{fontenc}
\usepackage{array}
\usepackage{makecell}
\usepackage{url}
\usepackage{lineno}
\usepackage{pifont}

\graphicspath{{figures/}}

\usepackage[breaklinks=true,bookmarks=false]{hyperref}

\iccvfinalcopy % *** Uncomment this line for the final submission

\def\iccvPaperID{****} % *** Enter the ICCV Paper ID here

\ificcvfinal\fi

\begin{document}
\title{Full-scale Representation Guided Network for Retinal Vessel Segmentation}

\author{Sunyong Seo\\
lululab, AI R\&D center\\
{\tt\small sy.seo@lulu-lab.com}
\and
Sangwook Yoo\\
lululab, AI R\&D center\\
{\tt\small sangwook.yoo@lulu-lab.com}
\and
Huisu Yoon\\
Department of Biomedical Engineering, University of Ulsan\\
{\tt\small hsyoon@ulsan.ac.kr}
}

\maketitle
\ificcvfinal\thispagestyle{empty}\fi

\begin{abstract}
The U-Net architecture and its variants have remained state-of-the-art (SOTA) for retinal vessel segmentation over the past decade. In this study, we introduce a Full-Scale Guided Network (FSG-Net), where a novel feature representation module using modernized convolution blocks effectively captures full-scale structural information, while a guided convolution block subsequently refines this information. Specifically, we introduce an attention-guided filter within the guided convolution block, leveraging its similarity to unsharp masking to enhance fine vascular structures. Passing full-scale information to the attention block facilitates the generation of more contextually relevant attention maps, which are then passed to the attention-guided filter, providing further refinement to the segmentation performance. The structure preceding the guided convolution block can be replaced by any U-Net variant, ensuring flexibility and scalability across various segmentation tasks. For a fair comparison, we re-implemented recent studies available in public repositories to evaluate their scalability and reproducibility. Our experiments demonstrate that, despite its compact architecture, FSG-Net delivers performance competitive with SOTA methods across multiple public datasets. Ablation studies further demonstrate that each proposed component meaningfully contributes to this competitive performance.
Our code is available on \url{https://github.com/ZombaSY/FSG-Net-pytorch}.
\end{abstract}

\section*{Introduction}\label{sect:intro}
Convolutional neural networks (CNNs) have seen significant improvements in performance and optimization since the 2010s. The introduction of hardware acceleration using GPUs, ReLU activation function, and residual block~\cite{He_2016_CVPR,https://doi.org/10.48550/arxiv.1802.06955} has enabled smooth back-propagation in deep neural network architectures. Research in this field has focused on finding a balance between computational efficiency, parameter size, code scalability, and fidelity. Depthwise separable convolution~\cite{Chollet_2017_CVPR} and squeeze-and-excitation~\cite{Hu_2018_CVPR} have been particularly influential in this regard. The inverted residual block~\cite{Sandler_2018_CVPR} achieved higher fidelity with optimized computational efficiency and a smaller parameter size compared to ResNet. 

\begin{figure}[t]
    \centering
    \includegraphics[width=0.8\linewidth]{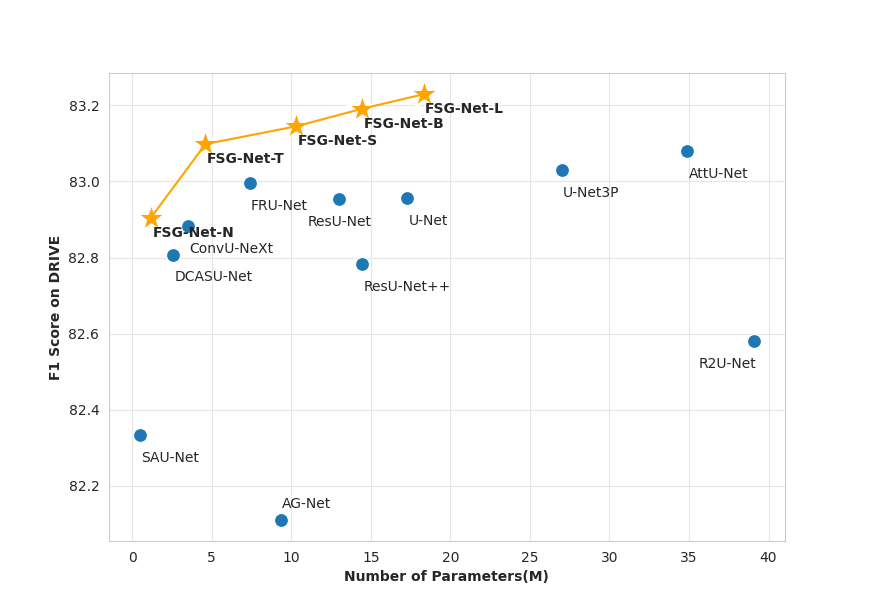}
    \caption{ \label{fig:F1_score_overview}F1 scores of compared networks on the DRIVE dataset, measured against the validation dataset comprising zero-padded images of resolution 608$\times$608. 
    Among the considered architectures, FSG-Net-T achieved a superior F1 score compared to competitive models while maintaining a reduced parameter size relative to its counterparts. Additionally, the FSG-Net achieved the highest F1 scores while possessing a median parameter size.}
\end{figure}

In the evolutionary history of CNNs, a noteworthy highlight is the dominance of U-Net~\cite{10.1007/978-3-319-24574-4_28} and its variants as SOTA models in the field of medical image segmentation across a wide spectrum of segmentation tasks~\cite{8959021, Tang_2022_CVPR, 9446143, 10.1007/978-3-030-00889-5_1, 9053405, 9413346, TIAN2020117, jin2022semi, jin2024inter, jin2021cascade, jin2021domain, tan2024deep, ding2024rcar}. In particular, retinal-vessel segmentation benefits from well-curated public datasets such as DRIVE, STARE, CHASE\_DB1, and HRF, on which the latest CNN-based methods are already systematically benchmarked and firmly established as strong baselines. In~\cite{qin2024review}, traditional and the recent deep learning based algorithms for retinal vessel segmentation are comprehensively reviewed.

On the contrary, for clinical segmentation tasks on the CVC-clinic and Kvasir-SEG datasets, vision transformer (ViT) models are employed~\cite{xie2021segformer, 10.1007/978-3-031-25066-8_9, chen2021transunet, HAN2022109512}. The primary distinction between retinal vessels and other clinical datasets lies in the level of feature intricacy required. 
ViT models exhibit some limitations in overcoming their constrained inductive bias. Meanwhile, our experiments underscore the ongoing relevance and effectiveness of attention mechanisms in addressing such challenges.

In this study, we aimed to propose a U-Net-based segmentation network that explicitly captures and preserves the thin and elongated structural characteristics of retinal vessels by leveraging a novel, full-scale feature representation strategy from the early layers upward. Furthermore, we introduced an attention-guided filtering approach inspired by guided filters~\cite{6319316}, strategically embedded in the decoder following encoder-decoder feature merging. This design allows the guided filter to perform adaptive edge enhancement, specifically utilizing comprehensive, full-scale structural information at each decoding stage. Our network architecture improves both performance and computational efficiency simultaneously, as shown in Fig.~\ref{fig:F1_score_overview}. To ensure a fair comparison of the robustness of competing models, the training environment was maintained consistently.

The main contributions of this paper are summarized as follows: Firstly, a novel convolutional block specifically designed for retinal vessel segmentation, capable of extracting comprehensive, full-scale structural features with computational efficiency. Secondly, an attention-guided convolutional block integrated into the decoder, enhancing segmentation accuracy through adaptive refinement, and offering flexibility to be combined seamlessly with various U-Net architectures. Thirdly, a thorough comparison and reproducibility study of multiple SOTA algorithms, conducted under strictly controlled experimental conditions.

\section*{Related work}\label{rel}
\subsection*{Guided filter}
The guided filter was first introduced in~\cite{6319316} for image processing under the assumption that the guidance image and the filtered output have a locally linear relationship. 
The guided filter is formulated as:
\begin{equation}\label{eq:gf_org}
\hat{I_{i}}=a_{k}I_{i}+b_{k},\quad\forall i\in \mathrm{w}_k,
\end{equation}
where $a_{k}$, $b_{k}$ are linear coefficients and assumed to be constants in the local window $\mathrm{w}_k$. $\hat{I}_{i}$ is the output of the guided filter with input $I_{i}$.
Here, we show why the guided filter can improve edge-like blood vessel segmentation performance. If we consider overlapping windows, the $i$-th pixel has several $a_k$ and $b_k$ values depending on the window size. Thus, the guided filter output of $I_i$  
can be averaged as follows:
\begin{equation}\label{eq:gf_aver}
\hat{I_{i}}=\frac{1}{|w|}\sum_{k\in{w_{i}}}\left(a_{k}I_{i}+b_{k}\right),\quad\forall i\in \mathrm{w}_k.
\end{equation}
The guided filter smooths images while preserving edge information, making it widely used in various image processing tasks such as image detail enhancement, image denoising, image preprocessing, and image dehazing.

\subsection*{Multiscale architectures preserving spatial detail}
In this study, we focus on two representative types of multiscale architectures for image segmentation: full-resolution preserving and full-scale aggregation approaches. In the full-resolution preserving category, HRNet maintains high-resolution representations by processing multi-resolution branches in parallel throughout the network~\cite{Sun_2019_CVPR}. Similarly, FR-UNet adopts a parallel full-resolution pathway while preserving the U-Net structure~\cite{9815506}. In the full-scale aggregation category, UNet3+ extends the skip-connection scheme of the original U-Net by enabling each decoder stage to simultaneously receive and fuse feature maps from all encoder and decoder stages~\cite{9053405}.

\subsection*{Deep supervision}
Deep supervision has recently become a prevalent strategy in various segmentation tasks, particularly within medical image analysis, to enhance model performance~\cite{10.1007/978-3-030-00889-5_1, wang2019deeply, reiss2021every}. In general, deep supervision refers to computing multiple loss functions at different intermediate layers of the neural network, enabling more stable gradient propagation throughout the training process. This approach has proven beneficial in reducing issues such as vanishing gradients, thus improving segmentation accuracy.

Several well-known variations of U-Net architectures, including M-Net~\cite{fu2018joint}, AG-Net~\cite{10.1007/978-3-030-32239-7_88}, and U-Net3+~\cite{9053405}, have successfully applied deep supervision and demonstrated better segmentation results compared to architectures utilizing a single loss function. Specifically, for retinal vessel segmentation, M-Net and AG-Net calculated multiple losses from each of the decoders, except the deepest one. On the other hand, U-Net3+ utilized losses from all decoder stages and employed them to train a segmentation model targeting liver and spleen structures in CT images.

\section*{Method}\label{sec2}

\subsection*{Motivation}

Now, (\ref{eq:gf_aver}) can be rewritten to have a more intuitive form. In the original paper of the guided filter~\cite{6319316}, $b_{k}$ is computed as:
\begin{equation}\label{eq:gf_b}
b_{k}=\bar{p}_{k}-a_{k}\mu_{k},
\end{equation} 
where $ \bar{\left(\cdot\right)}_{k} $ means average in a window $\mathrm{w}_k$ and $\mu_{k}$ is the mean of $I_{k}$. Then, by putting Eq.~(\ref{eq:gf_b}) into Eq.~(\ref{eq:gf_aver}), we have the following formulation:  
\begin{equation}\label{eq:umgf}
\hat{I}_{i}=\bar{a}_i(I_{i}-\mu_{i})+\tilde{p}_{i},
\end{equation}
where $\tilde{\left(\cdot\right)}_{k}$ denotes the average of the average, meaning $\tilde{p}_{i}=\frac{1}{|w|}\sum_{k \in w_{i}}{\bar{p}}_{k}$. The unsharp masking is defined as follows:
%$\hat{I}=\alpha\left(I-L(I)\right)+I$, 
\begin{equation}\label{eq:UM}
\hat{I}_i = \alpha \left( I_i - \sum_{k \in \mathrm{w}_i} w_{ik} I_k \right) + I_i,
\end{equation}
where $w_{ik}$ denotes the normalized Gaussian weight centered at pixel $i$, evaluated at neighboring pixel $k$. 
%where $L(\cdot)$ is a low-pass filter such as the Gaussian blur.
Eq.~(\ref{eq:umgf}) looks like Eq.~(\ref{eq:UM}), where sharpening mask $\left(I_{i}-\bar{I}_{i}\right)$ from the guidance image is added to the averaged target image. The filter strength is controlled by $\alpha$.

In their pioneering work~\cite{10.1007/978-3-030-32239-7_88}, Zhang et al. adopted the guided filter in the segmentation network and suggested incorporating the attention map $M$ into the energy minimization problem to estimate $a$ and $b$:

\begin{equation}\label{eq:qf2}
\mathcal{E}(a_k, b_k) = \sum_{i \in \mathrm{w}_k} \left( M_i^2 \left( a_k {I_d}_i + b_k - g_i \right)^2 + \epsilon a_k^2 \right).
\end{equation}

In Eq.~(\ref{eq:qf2}), the ${I_{d}}_{i}$ is the downsampled input feature from encoding parts, where $g_{i}$ denotes the gating signal from
upward path at the decoding parts.
Eq.~(\ref{eq:qf2}) implies that an improved attention map can lead to better solutions. In the following, we are going to propose a method that can generate an improved attention map with full-scale information.

\subsection*{Network architecture}

\begin{figure*}[t]
    \centering
    \includegraphics[width=0.9\linewidth]{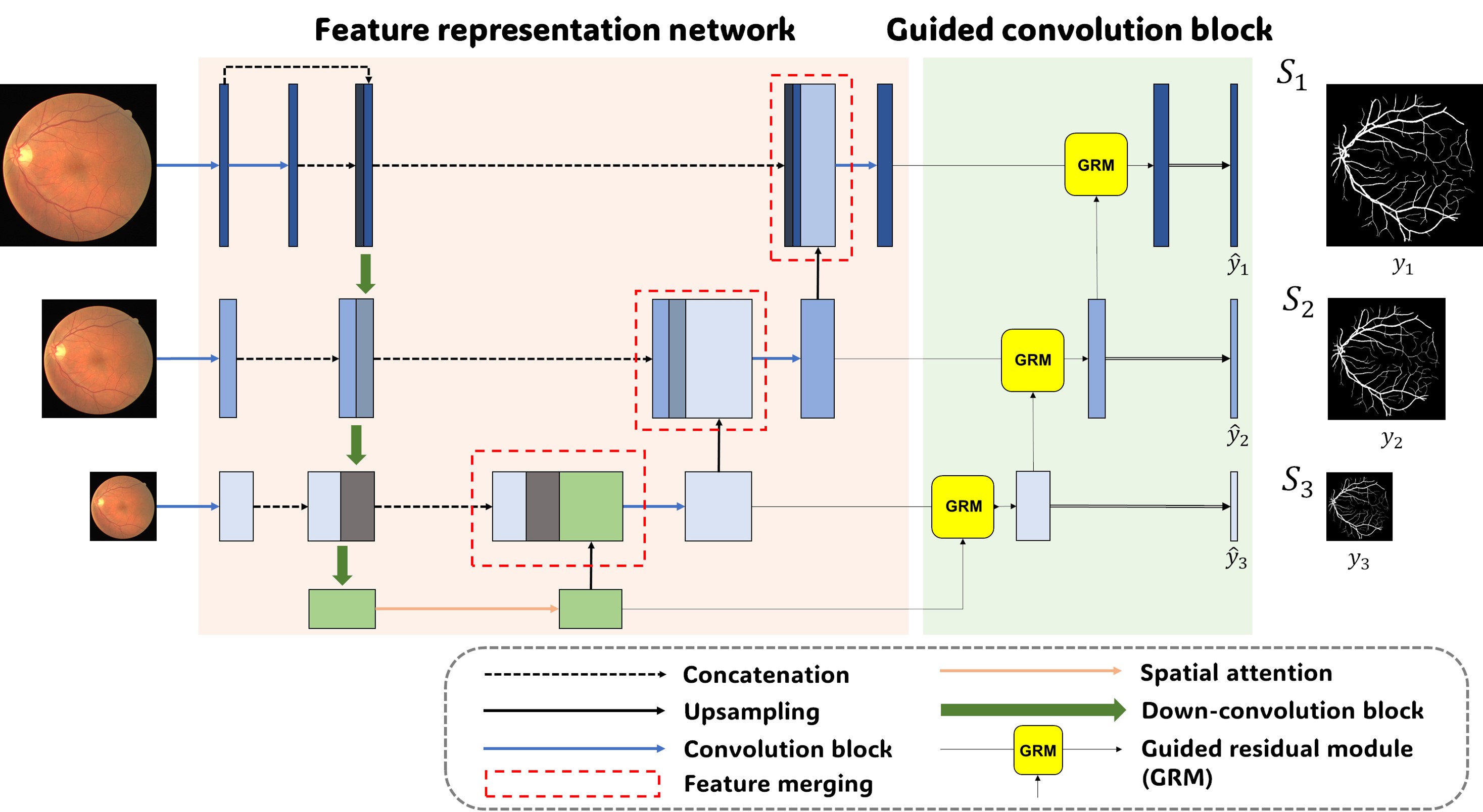}
    \caption{
        \label{fig:neU-Net_overview}
           Network architecture of the proposed FSG-Net. In the feature representation network (left), down-convolution layers concatenate separately extracted features, connected via deep bottleneck structures to up-convolution layers. The red dashed boxes highlight feature merging from multiple scales. The guided convolution block (right) employs Guided Residual Modules (GRM) to refine multi-scale features. $S_i$ indicates each stage, and $y_i$ represents labels, where $y_2$ and $y_3$ are downsampled from the original label $y_1$ by interpolation.
           }
\end{figure*}
As can be seen in Fig.~\ref{fig:neU-Net_overview}, the proposed network basically follows the U-Net architecture but has an additional guided convolution block after the feature representation network, unlike usual U-Net based structures. 
Contrary to the standard U-Net consisting of five stages, we have opted for four stages in FSG-Net based on the assumption that a wider receptive field is less critical for retinal vessel segmentation.

As illustrated on the left side of the FSG-Net architecture, the input to each down-convolution block, which is indicated by the bold green arrows in the figure, is augmented by concatenating features derived from an auxiliary convolutional path. These preserved features are subsequently forwarded to the corresponding up-convolution layers via skip connections, facilitating richer multi-scale feature reuse. The convolutional blocks within this stage adopt a deep bottleneck structure, which enhances the network's representational capacity while maintaining computational efficiency. Furthermore, a lightweight spatial attention, inspired by the Convolutional Block Attention Module (CBAM)~\cite{Woo_2018_ECCV}, is integrated into the bottleneck stage to refine spatial feature encoding with minimal parameter overhead.

In the feature representation network, the key area to note is the feature merging section, indicated by the red dashed box. In this feature merging process, features from three paths—the current, upper, and lower stages—are concatenated and passed through a newly designed convolution block before being forwarded to the subsequent stages. This approach allows the integration and transmission of information across all scales. In this study, we refer to such integrated multi-scale information as full-scale information. The advantages of gathering the entire information in this manner will be further discussed in the Results section.

In the right part of the model, namely the guided convolution block, compressed features reflecting full scale information are given as the current stage input to the guided residual module (GRM). Up-stage input from the higher stage is given as another input for guided filtering. After GRM, predictions are derived through convolution and activation.
For training, a combination of the BCE loss and Dice loss with deep supervision was employed to improve 
segmentation performance of the edge-like structures~\cite{rs15061530}:

\begin{equation}\label{eq:DS}
L_{\text{DS}} = \sum_{d=1}^{D} \alpha_d \cdot \left( L_\text{BCE}^{(d)} + \lambda \cdot L_\text{Dice}^{(d)} \right),
\end{equation}
\begin{equation}\label{eq:BCE}
\text{where} \quad
L_\text{BCE}^{(d)} = -\frac{1}{S_{d}} \sum_{i=1}^{S_{d}} \left( y_{i,d} \log \hat{y}_{i,d} + (1 - y_{i,d}) \log (1 - \hat{y}_{i,d} \right),
\end{equation}
\begin{equation}\label{eq:Dice}
\text{and} \quad
L_\text{Dice}^{(d)} = 1 - \frac{2 \sum_{i=1}^{S_{d}} y_{i,d} \hat{y}_{i,d} + \epsilon}{\sum_{i=1}^{S_{d}} y_{i,d} + \sum_{i=1}^{S_{d}} \hat{y}_{i,d} + \epsilon}.
\end{equation}

In Eq.~(\ref{eq:DS}), $D$, $\alpha_d$ represent the number of prediction layers and weights for each prediction layer $d$, respectively.
In Eq.~(\ref{eq:BCE}) and Eq.~(\ref{eq:Dice}), $S_{d}$ denotes the number of samples at the $d$-th prediction layer and $y_{i,d}$ and $\hat{y}_{i,d}$ represent target and prediction values at $i$-th index and at $d$-th prediction level, respectively.  

Lastly, we can consider the scalability of the proposed architecture. 
The feature representation network can be replaced by other U-Net variants, enabling integration with the guided convolution block. 
For instance, we can consider integrating with FR-UNet~\cite{9815506}, which proposes alternative methods for utilizing full-scale information.

\subsection*{Modernized convolution block}
The standard U-Net has a fundamental structure comprising skip connections and double convolution blocks. Recent studies have shown that incorporating different CNN structures can significantly improve the performance of the original model~\cite{liu2022convnet, HAN2022109512}. Inspired by the ConvNeXt~\cite{liu2022convnet}, we have designed a convolutional block suitable for retinal vessel segmentation.

\begin{figure*}[htp]
    \centering
    \includegraphics[width=0.85\linewidth]{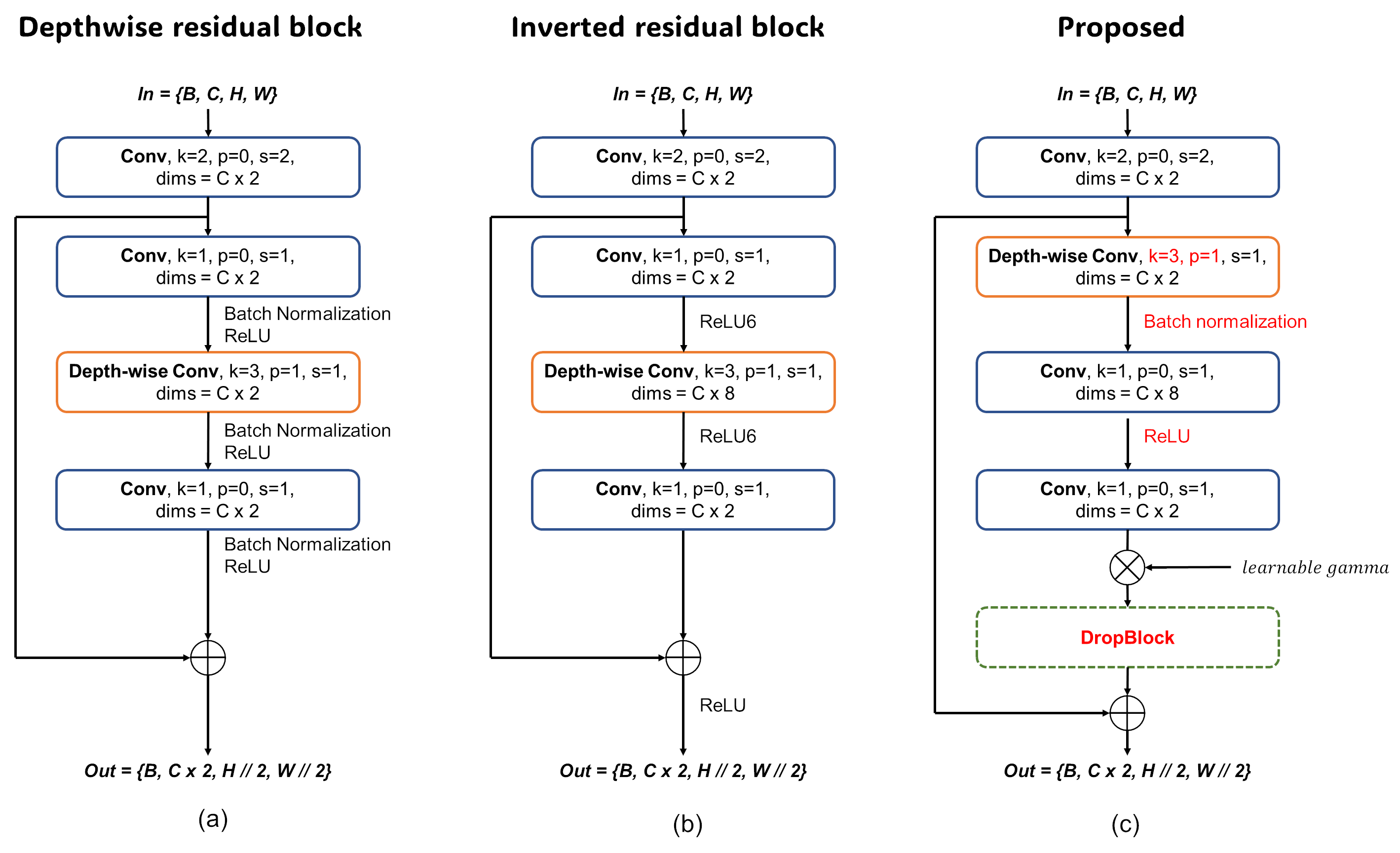}
    \caption{
        \label{fig:up_down_conv}
           The evolutionary structure from (a): Depthwise residual block, (b): Inverted residual block to (c): The proposed convolution block.
           }
\end{figure*}

Fig.~\ref{fig:up_down_conv}(a) and (b) represent the depthwise residual block and inverted residual block, respectively. 
Fig.~\ref{fig:up_down_conv}(c) shows the proposed convolution block. The structure of the block is an extension of the latest advancements in convolutional block development, characterized by features such as 1\(\times\)1 convolution, inverted bottleneck, and depth-wise convolution. Like other modernized convolution blocks, the proposed block employs a convolution with kernel size 2 and stride 2 in the first stage. By incorporating spatial and dimensional changes within the first block, we enable the design of a deep bottleneck stage in the down-convolution structure. This approach not only increases the information on the feature but also separates it from the bottleneck, allowing for more detailed feature representation. Furthermore, to maintain linearity, we utilize unique ReLU activation between each inverted residual block. For regularization purposes, we employ a learnable gamma parameter and apply a drop block before joining the identity block.

\begin{figure*}[htp]
    \centering
    \includegraphics[width=0.85\linewidth]{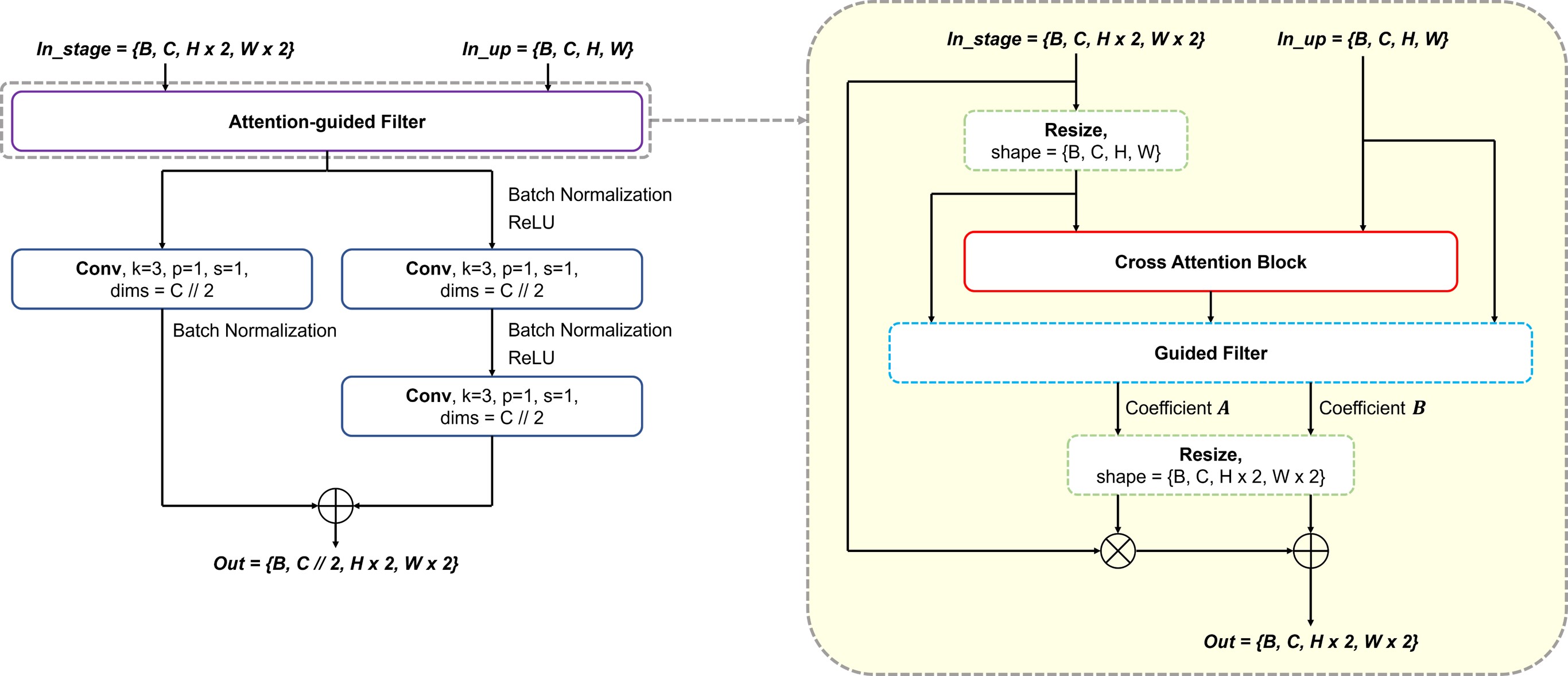}
    \caption{
        \label{fig:attention_guided_filter}
           Detailed structure of guided residual module (GRM)
           }
\end{figure*}

\section*{Experiments}

\subsection*{Guided convolution block}
As shown in Fig.~\ref{fig:neU-Net_overview}, the guided convolution block consists of GRM to refine the input from the feature representation network and a convolution to output the prediction map. 
Fig.~\ref{fig:attention_guided_filter} shows the overall process of GRM. 

The purpose of GRM is to refine the feature from the feature representation network at the same stage.
In GRM, both the features at the current stage and at the higher stage are used to generate an attention map, and then the attention map is multiplied with the current feature to generate the output, which is described in the right part of Fig.~\ref{fig:attention_guided_filter}. Here we can expect an improved attention map $M$ can be estimated to solve Eq.~(\ref{eq:qf2}) because the input feature from the feature representation network has full scale information.  
To further enhance the feature refinement, a residual block is introduced after the attention-guided filtering, as shown in the left part of Fig.~\ref{fig:attention_guided_filter}. By passing the features through the residual block, a stable map is generated for both the deep supervision and subsequent layers. Moreover, we incorporate a $1\times1$ convolution to preserve the semantic information present in the feature maps.

\subsection*{Training Techniques}
In retinal vessel segmentation, the performance of studies is often determined by subtle gaps. We believe that these subtle gaps are highly influenced by the choice of hyperparameters and training/inference environment. To address this imbalance, we fixed all hyperparameters for training and inference. We empirically found that RandAugment~\cite{Cubuk_2020_CVPR_Workshops} with a specific scale did not work well on medical datasets; therefore, we customized it to better suit our datasets. Training techniques include blur, color jitter, horizontal flip, perspective transformation, resize, crop, and CutMix~\cite{Yun_2019_ICCV}.

\subsection*{Implementation Details}
Our experimental environment comprises an Intel Xeon Gold 5220 processor, a Tesla V100-SXM2-32GB GPU, Pytorch 1.13.1, and CUDA version 11.7. The inference time for FSG-Net was approximately \(600\,\mathrm{ms}\) for an input size of 608$\times$608, where the original DRIVE image size 565$\times$584 was zero-padded to the nearest multiple of 32 for compatibility with the network structure.
To address as much variability as possible, we re-implemented comparison studies and integrated them into a single environment. To ensure experimental fairness, certain hyperparameters, including the framework, loss function, metric, data augmentation, and random seed, were fixed to measure the robustness of the model. Our training recipe followed the hyperparameters in Table ~\ref{tab:hyper-parameters}, used for segmentation tasks in ADE20K multiscale learning in ConvNeXt~\cite{liu2022convnet} as our proposed down-convolution module is directly derived from the ConvNeXt block.

\begin{table*}[h]
    \renewcommand{\arraystretch}{1.15}
    \centering
    \caption{\label{tab:hyper-parameters}
        Train settings and hyper-parameters.
    }
    \scriptsize
    \begin{tabular}{ l | c c}
        \Xhline{2\arrayrulewidth}
        \makecell{\textbf{Hyper-parameters}} & \makecell{\textbf{Values}} \\
        \hline
        \hline
        base lr & 1e-3 \\
        lr scheduler &  Linear warm-up, Cosine annealing\\
        lr scheduler warm-up epochs & 20 \\
        lr scheduler cycle epochs & 100 \\
        lr scheduler eta min & 1e-5 \\
        early stop epochs & 400 \\
        early stop metric & F1 score \\
        optimizer & AdamW \\
        optimizer momentum & \(\beta_1, \beta_2\)=0.9, 0.999 \\
        weight decay & 0.05 \\
        criterion & Dice + BCE \\
        binary threshold & 0.5 \\
        batch size & 4 \\
        center padded shape (dataset) & (D=608, S=704, C=1024, H=1344) \\
        random crop & 288 \\
        random blur & gaussian, k$\in$\{3, 5, 7, 9, 11\}, prob=0.8 \\
        random jitter & b=0.2, c=0.2, s=0.2, h=0.1, prob=0.8 \\
        random horizontal flip & prob=0.5 \\
        random perspective & s=0.3, prob=0.3 \\
        random random resize & s=[0.5, 2.0], prob=0.8 \\
        CutMix & n=1, prob=0.8 \\
        \Xhline{2\arrayrulewidth}
    \end{tabular}
\end{table*}

To evaluate the compared models under the same conditions, we prioritized the search for an optimized model in our training settings. For example, the learning rate can affect the gradient updating and training time, depending on the model's parameter size and depth. Training for a predetermined number of epochs can result in diverging weights for heavy models and, conversely, for light models. Therefore, we chose the optimized model using an early stop based on cycles in the learning rate scheduler. To select the optimized model during the training step, we used the highest F1 score~\cite{8309343}, with an early stop of 400 epochs. To balance exploitation and exploration in the learning parameters, we stack the batch to have more than two sets of mini-batches in one epoch with a learning rate scheduler. The detailed hyper-parameters are described in Table \ref{tab:hyper-parameters}. With these experimental settings, the performance of the pure U-Net dramatically increased and even surpassed that of some recent studies, as shown in Table \ref{tab:Comparison_Metrics}.

\subsection*{Datasets}
The DRIVE dataset comprised 40 retinal images with a resolution of 565\(\times\)584 pixels, captured as part of a retinopathy screening study in the Netherlands. The STARE dataset comprises 20 retinal fundus images with a resolution of 700 × 605 pixels, and the CHASE\_DB1 dataset includes 28 retinal images from schoolchildren with a resolution of 999\(\times\)960 pixels. Both the STARE and CHASE\_DB1 datasets were manually annotated by two independent experts. We used the annotation of the first expert, named "Hoover A." in STARE and "1stHO" in CHASE\_DB1, for our analysis. The HRF dataset comprises 45 images, equally divided into a 1:1:1 ratio of healthy patients, diabetic retinopaths, and glaucomatous patients, with a high resolution of 3504\(\times\)2336 pixels. To measure the performance of the models, it is necessary to divide the data into training and validation sets. As the retinal vessel segmentation dataset was relatively limited, we split the data into a 1:1 ratio of the training and validation sets. The DRIVE dataset was officially divided into training and validation sets, each containing 20 images. For the STARE, CHASE\_DB1, and HRF datasets, we used the first half as training and the remaining half as validation. 

\begin{table*}[htp]
    \renewcommand{\arraystretch}{0.8}
    \centering
    \caption{\label{tab:Comparison_Metrics}
        Comparison of segmentation performance for CNN-based networks.
    }
    \resizebox{0.95\textwidth}{!}{%
    \begin{tabular}{ l l | c c c c c c c c c}
        \Xhline{2\arrayrulewidth}
        \makecell{} & \makecell{\textbf{Architecture}} & \makecell{\textbf{mIoU}} & \makecell{\textbf{F1 score}} & \makecell{\textbf{Acc}} & \makecell{\textbf{AUC}} &  \makecell{\textbf{Sen}} & \makecell{\textbf{MCC}} & \makecell{\textbf{Rank Avg}} \\
        \hline
        \hline
        \noalign{\vskip 1pt}
        \textbf{DRIVE} & \\
        & U-Net & 83.857 & 82.956 & 97.013 & 97.853 & 83.449 & 81.456 & 5.7 \\
        & U-Net++ & 81.228 & 79.564 & 96.524 & 96.271 & 77.802 & 77.830 & 14.0 \\
        & U-Net3+ Deep & 83.909 & 83.030 & 97.017 & 98.082 & 83.721 & 81.520 & 3.5 \\
        & ResU-Net & 83.862 & 82.953 & 97.021 & 97.766 & 83.226 & 81.453 & 6.5 \\
        & ResU-Net++ & 83.729 & 82.783 & 97.001 & 97.708 & 82.791 & 81.263 & 10.2 \\
        & SAU-Net & 83.368 & 82.334 & 96.925 & 97.616 & 82.311 & 80.782 & 12.2 \\
        & DCASU-Net & 83.743 & 82.808 & 96.996 & 97.838 & 83.080 & 81.290 & 8.7 \\
        & AG-Net & 83.176 & 82.111 & 96.882 & 97.628 & 82.155 & 80.540 & 12.8 \\
        & AttU-Net & 83.958 & 83.080 & 97.039 & 97.844 & 83.422 & 81.584 & 3.5 \\
        & R2U-Net & 83.555 & 82.580 & 96.952 & 97.879 & 82.961 & 81.038 & 9.7 \\
        & ConvU-NeXt & 83.800 & 82.882 & 97.012 & 97.835 & 83.019 & 81.367 & 7.8 \\
        & FR-UNet & 83.884 & 82.995 & 97.007 & 98.158 & 83.869 & 81.485 & 4.3 \\
        & HRNet & 83.938 & 83.829 & \textbf{97.325} & 97.860 & 82.963 & 81.506 & 5.0 \\
        & \textbf{FSG-Net (ours)} & \textbf{84.068} & \textbf{83.229} & 97.042 & \textbf{98.235} & \textbf{84.207} & \textbf{81.731} & \textbf{1.2} \\
        \hline
        \noalign{\vskip 1pt}
        \textbf{STARE} & & & & & & & & & \\ 

        & U-Net & 85.924 & 84.873 & 97.754 & 98.341 & 84.361 & 83.713 & 3.7 \\
        & U-Net++ & 81.514 & 79.061 & 97.022 & 94.479 & 75.539 & 77.764 & 13.8 \\
        & U-Net3+ Deep & 85.824 & 84.829 & 97.707 & \textbf{99.146} & 85.522 & 83.626 & 3.7 \\
        & ResU-Net & 85.964 & 84.872 & 97.767 & 98.050 & 83.997 & 83.726 & 4.0 \\
        & ResU-Net++ & 83.185 & 81.358 & 97.319 & 95.221 & 78.456 & 80.196 & 12.7 \\
        & SAU-Net & 85.158 & 84.061 & 97.604 & 97.784 & 84.015 & 82.885 & 8.0 \\
        & DCASU-Net & 84.423 & 83.064 & 97.454 & 98.362 & 83.583 & 81.771 & 9.7 \\
        & AG-Net & 84.811 & 83.766 & 97.565 & 98.403 & 83.347 & 82.516 & 8.5 \\
        & AttU-Net & 85.848 & 84.772 & 97.694 & 99.050 & 86.226 & 83.588 & 4.0 \\
        & R2U-Net & 83.727 & 81.786 & 97.468 & 93.457 & 77.048 & 80.810 & 12.3 \\
        & ConvU-NeXt & 85.339 & 84.186 & 97.658 & 97.866 & 83.401 & 82.998 & 7.5 \\
        & FR-UNet & 84.815 & 83.496 & 97.577 & 96.637 & 81.872 & 82.327 & 10.2 \\
        & HRNet & 85.555 & 84.165 & \textbf{97.982} & 98.190 & 84.604 & 83.197 & 5.2 \\
        & \textbf{FSG-Net (ours)} & \textbf{86.118} & \textbf{85.100} & 97.746 & 98.967 & \textbf{86.608} & \textbf{83.958} & \textbf{1.8} \\
        \hline
        \noalign{\vskip 1pt}
        \textbf{CHASE\_DB1} & & & & & & & & & \\

        & U-Net & 82.065 & 80.159 & 97.404 & 99.368 & 85.370 & 79.002 & 7.0 \\
        & U-Net++ & 81.512 & 79.415 & 97.321 & 99.362 & 84.143 & 78.201 & 9.2 \\
        & U-Net3+ Deep & 82.489 & 80.697 & 97.483 & 99.506 & 85.740 & 79.558 & 3.7 \\
        & ResU-Net & 81.104 & 78.810 & 97.320 & 99.278 & 81.311 & 77.479 & 12.5 \\
        & ResU-Net++ & 73.276 & 67.411 & 95.966 & 96.223 & 67.782 & 65.593 & 14.0 \\
        & SAU-Net & 81.335 & 79.100 & 97.338 & 99.407 & 82.265 & 77.802 & 9.8 \\
        & DCASU-Net & 82.254 & 80.368 & 97.483 & 99.329 & 83.916 & 79.161 & 6.3 \\
        & AG-Net & 82.158 & 80.272 & 97.440 & 99.544 & 84.817 & 79.070 & 5.2 \\
        & AttU-Net & 82.562 & 80.742 & 97.546 & 99.430 & 83.907 & 79.537 & 4.3 \\
        & R2U-Net & 81.250 & 78.944 & 97.384 & 98.792 & 80.054 & 77.625 & 11.8 \\
        & ConvU-NeXt & 81.752 & 79.704 & 97.405 & 99.402 & 83.009 & 78.439 & 8 \\
        & FR-UNet & 81.330 & 79.170 & 97.269 & 99.544 & 84.744 & 77.944 & 8.7 \\
        & HRNet & \textbf{82.849} & \textbf{81.021} & \textbf{97.744} & \textbf{99.650} & 85.750 & \textbf{79.987} & 1.2 \\
        & \textbf{FSG-Net (ours)} & 82.680 & 81.019 & 97.515 & 99.378 & \textbf{85.995} & 79.899 & \textbf{3.0} \\

        \hline
        \noalign{\vskip 1pt}
        \textbf{HRF} & & & & & & & & & \\

        & U-Net & 82.291 & 81.329 & 97.093 & 98.571 & 82.756 & 79.868 & 4.8 \\
        & U-Net++ & 82.653 & 80.995 & 97.040 & 98.450 & 82.402 & 79.522 & 7.8 \\
        & U-Net3+ Deep & 83.006 & 81.445 & \textbf{97.124} & 98.427 & 82.436 & 79.997 & 4.4 \\
        & ResU-Net & 82.908 & 81.299 & 97.098 & 98.415 & 82.481 & 79.842 & 7.0 \\
        & ResU-Net++ & 77.008 & 73.096 & 96.032 & 95.312 & 70.587 & 71.251 & 13.8 \\
        & SAU-Net & 82.015 & 80.170 & 96.892 & 98.539 & 82.195 & 78.624 & 10.3 \\
        & DCASU-Net & 82.942 & 71.335 & 97.109 & 98.499 & 82.383 & 79.884 & 7.3 \\
        & AG-Net & 82.249 & 80.474 & 96.961 & 98.566 & 81.810 & 78.951 & 8.7 \\
        & AttU-Net & 83.017 & 81.448 & 97.113 & 98.558 & 82.830 & 79.996 & 2.8 \\
        & R2U-Net & 80.998 & 78.699 & 96.808 & 96.976 & 77.231 & 77.122 & 12.8 \\
        & ConvU-NeXt & 83.015 & 81.423 & 97.121 & 98.525 & 82.520 & 79.974 & 4.0 \\
        & FR-UNet & 82.431 & 80.709 & 97.011 & 98.294 & 81.677 & 79.218 & 9.7 \\
        & HRNet & 82.067 & 80.226 & 96.944 & 98.563 & 80.957 & 78.668 & 9.8 \\
        & \textbf{FSG-Net (ours)} & \textbf{83.088} & \textbf{81.567} & 97.106 & \textbf{98.744} & \textbf{83.616} & \textbf{80.121} & \textbf{1.6} \\
        \Xhline{2\arrayrulewidth}
    \end{tabular}
    }
\end{table*}

\subsection*{Results}
In binary task evaluations, the Matthew correlation coefficient (MCC) is a powerful metric, as noted by Chicco~\textit{et al.}~\cite{Chicco2021TheBO}. However, to avoid evaluations oriented towards a specific metric, we also report the average rank of each model, denoted as "Rank Avg" in Table \ref{tab:Comparison_Metrics}. This average rank provides a measure of the stable performance of a model across different datasets. For example, FSG-Net, U-Net3+~\cite{9053405}, and AttU-Net~\cite{DBLP:journals/corr/abs-1804-03999} achieved high ranks in all four datasets, whereas ResU-Net, FR-UNet and HRNet~\cite{Sun_2019_CVPR} recorded inconsistent results across the three datasets. FSG-Net consistently demonstrated top-tier performance across all four datasets, recording dominant scores in mIoU, F1 score, and MCC, which is equivalent to a detailed expression of the segmentation map. Notably, FSG-Net outperformed previous methods on the DRIVE dataset, achieving SOTA performance in F1 score and sensitivity.

We conducted comparative experiments with existing architectures that aim to preserve full-resolution features, namely HRNet and FR-UNet.
As shown in Table~\ref{tab:Comparison_Metrics}, HRNet achieves the best performance on the CHASE\_DB1 dataset and also records the highest accuracy on the DRIVE and STARE datasets. However, its performance noticeably drops on the HRF dataset, and notably, HRNet contains 65.81M parameters, which significantly exceeds the model capacities of all other compared architectures that maintain parameter counts below 30M. FR-UNet performs well on the DRIVE dataset but shows relatively moderate results on the remaining datasets.
These comparisons with high-resolution preserving models demonstrate the performance stability and robustness of FSG-Net across diverse datasets, validating the effectiveness of utilizing full-scale information at multiple decoding stages.

\begin{table*}[htp]
    \centering
    \tiny
    \caption{\label{tab:ViT_metric}
        Performance comparison of ViT-based models and FSG-Net on four retinal vessel segmentation datasets.
    }
    \resizebox{0.9\textwidth}{!}{%
    \begin{tabular}{ l l | c c c c c c c c}
        \Xhline{2\arrayrulewidth}
        \makecell{} & \makecell{\textbf{Architecture}} & \makecell{\textbf{mIoU}} & \makecell{\textbf{F1 score}} & \makecell{\textbf{Acc}} & \makecell{\textbf{AUC}} &  \makecell{\textbf{Sen}} & \makecell{\textbf{MCC}} \\
        
        \hline
        \hline
        \noalign{\vskip 1pt}
        \textbf{DRIVE} & \\
        & Swin & 78.758 & 76.040 & 96.296 & 97.203 & 75.681 & 74.171 \\
        & MaxViT-512 & 79.702 & 77.738 & 96.050 & 98.452 & 79.097 & 75.693 \\
        & FSG-Net (ours) & 84.068 & 83.229 & 97.042 & 98.235 & 84.207 & 81.731 \\

        \hline
        \noalign{\vskip 1pt}
        \textbf{STARE} & & & & & & & & & \\
        & Swin& 79.354 & 75.900 & 96.958 & 97.759 & 75.186 & 74.356 \\
        & MaxViT-512 & 81.072 & 78.898 & 96.816 & 97.865 & 79.667 & 77.357 \\
        & FSG-Net (ours) & 86.118 & 85.100 & 97.746 & 98.967 & 86.608 & 83.958 \\

        \hline
        \noalign{\vskip 1pt}
        \textbf{CHASE\_DB1} & & & & & & & & & \\
        & Swin & 77.708 & 73.850 & 96.966 & 99.326 & 76.569 & 72.347 \\
        & MaxViT-512 & 79.322 & 76.405 & 96.978 & 99.494 & 79.878 & 74.909 \\
        & FSG-Net (ours) & 82.680 & 81.019 & 97.515 & 99.378 & 85.995 & 79.899 \\

        \hline
        \noalign{\vskip 1pt}
        \textbf{HRF} & & & & & & & & & \\
        & Swin & 78.739 & 74.771 & 97.420 & 98.022 & 74.831 & 73.548 \\
        & MaxViT-512 & 74.980 & 70.617 & 95.346 & 98.645 & 72.600 & 68.204 \\
        & FSG-Net (ours) & 83.088 & 81.567 & 97.106 & 98.744 & 83.616 & 80.121 \\
        
        \Xhline{2\arrayrulewidth}
    \end{tabular}
    }
\end{table*}

We have also conducted experiments to assess the suitability of ViT-based models for the task of retinal vessel segmentation. As noted in the introduction, the thin and elongated structure of retinal vessels poses specific challenges, particularly for models that rely heavily on global representations such as pure ViT. To explore this further, we evaluated Swin-T~\cite{Liu_2021_ICCV}, which introduces a hierarchical representation with shifted window attention, and MaxViT-T-512~\cite{10.1007/978-3-031-20053-3_27}, a hybrid model that combines convolution and attention mechanisms and was evaluated using the UPerNet~\cite{Xiao_2018_ECCV} decoding head.
As shown in Table~\ref{tab:ViT_metric}, the ViT-based models in our setting yielded lower overall performance compared to FSG-Net, particularly in metrics such as F1 score and sensitivity. Swin-T and MaxViT-T-512 contain 58.91M and 59.60M parameters, respectively, yet FSG-Net achieves substantially better performance despite operating with a more parameter-efficient design.

Nonetheless, recent advances in DETR-like architectures~\cite{10.1007/978-3-030-58452-8_13} have introduced object query mechanisms that demonstrate strong potential for enhancing small object detection. These efforts reflect ongoing attempts to overcome the limitations traditionally associated with representing fine, low-saliency structures using transformer-based models. Although not included in our experiments, TCDDU-Net~\cite{lv2024tcddu}, which combines a dual-path U-Net architecture with a Swin backbone, has been reported to outperform several CNN-based methods in terms of quantitative metrics and to effectively segment peripheral vessels. Although not directly comparable due to different experimental settings, it is notable that TCDDU-Net achieved an F1 score of 82.65 on the DRIVE dataset, which is slightly lower than the 84.068 recorded by our proposed FSG-Net. These research findings suggest that future research may benefit from further exploring hybrid architectures or improving transformer-based designs to better handle fine-grained and low-saliency structures such as retinal vessels.

\begin{figure*}[htp]
    \centering
    \includegraphics[width=0.8\linewidth]{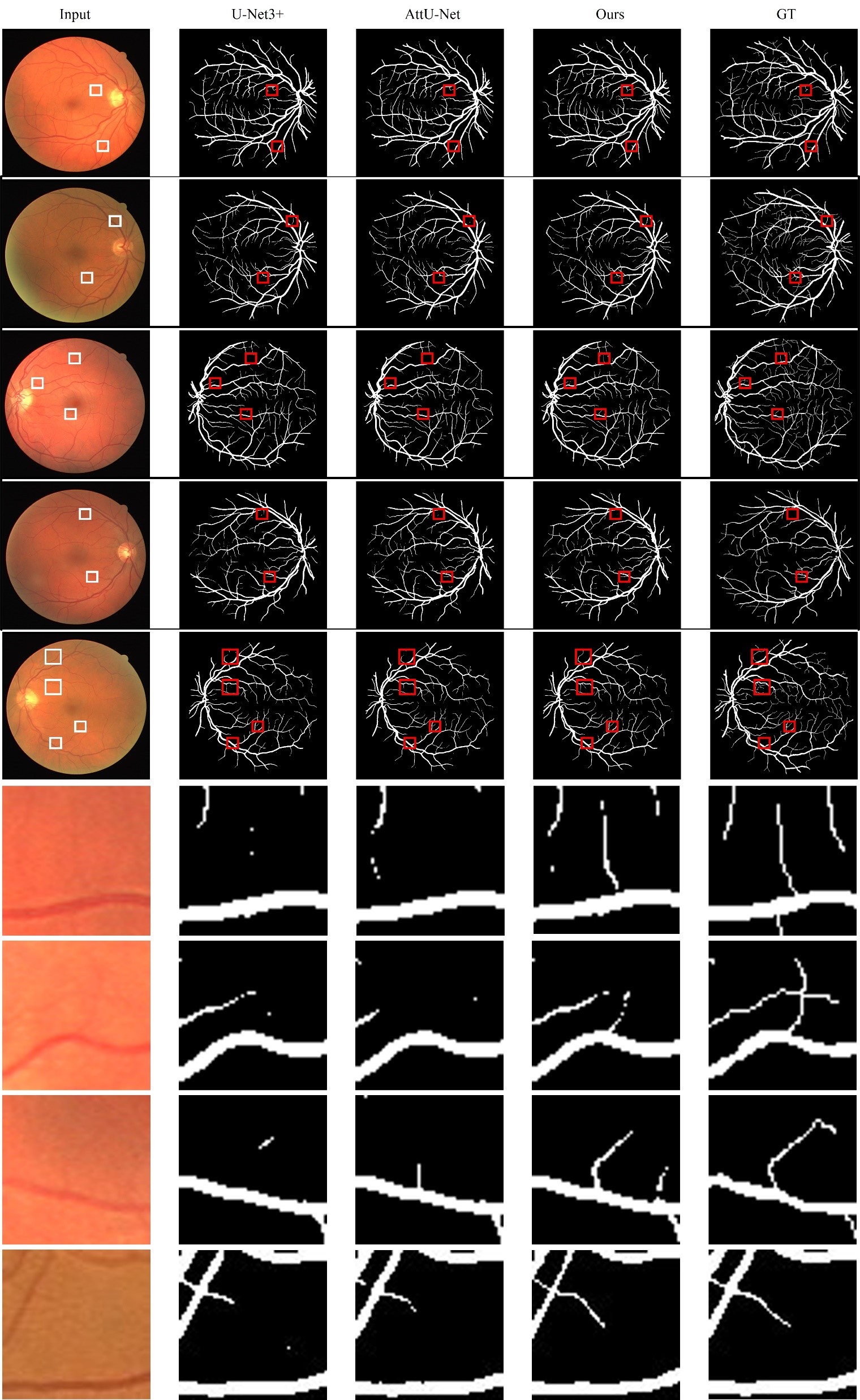}
    \caption{\label{fig:Qualitative_evaluation}
          Qualitative comparison of the top-3 performing models on the DRIVE validation set.
          }
\end{figure*}

\begin{figure*}[htp]
    \centering
    \includegraphics[width=0.8\linewidth]{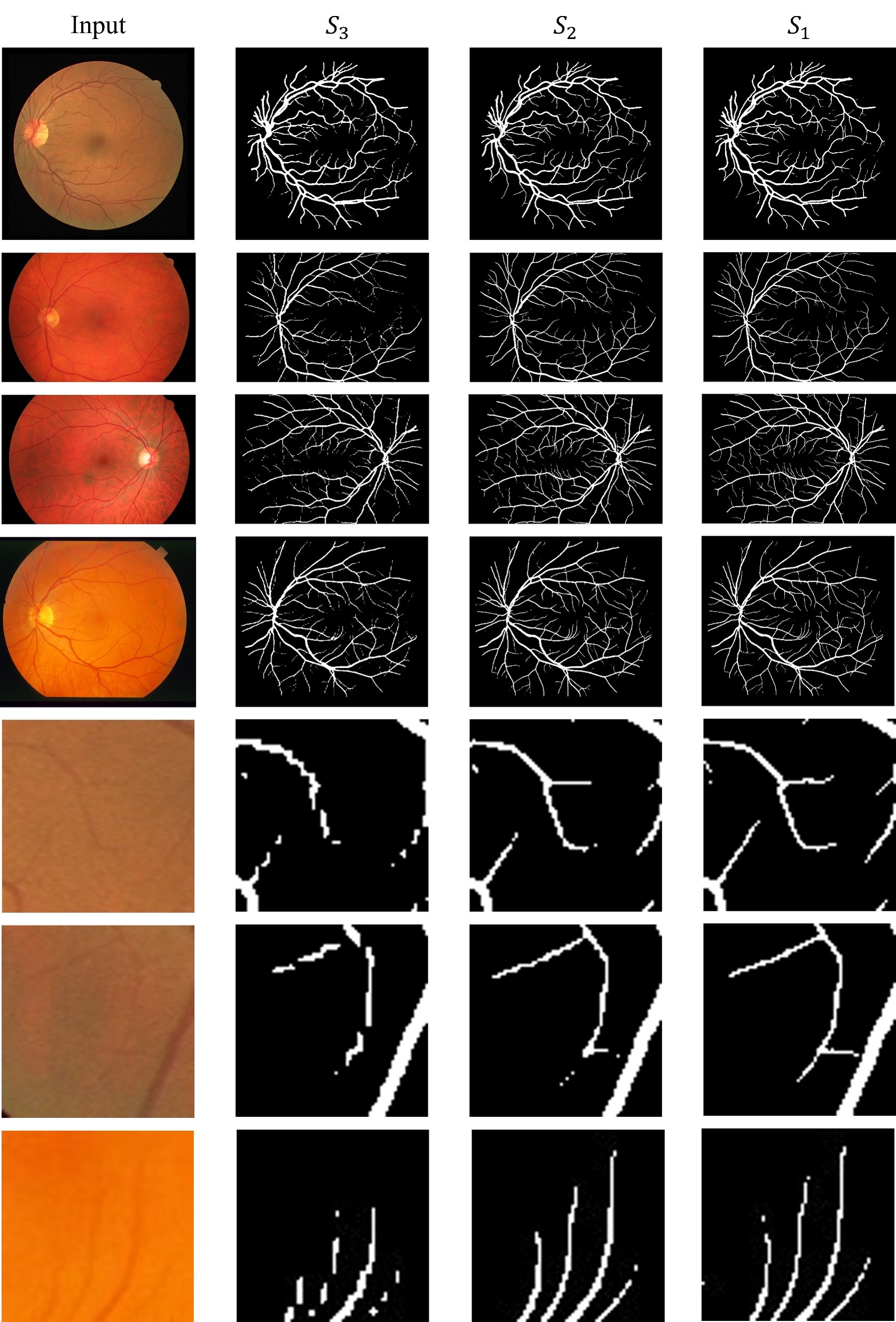}
    \caption{\label{fig:Stage_Comprison}
          Visualization of predictions at each stage $S_i$ of FSG-Net. Here, $S_i$ represents each decoder stage, and a larger value of $i$ indicates a deeper decoder as shown in Fig.~\ref{fig:neU-Net_overview}. The first four rows show the entire prediction results from various datasets, while the last three rows present magnified views of these predictions.
          }
\end{figure*}

Fig.~\ref{fig:Qualitative_evaluation} represents the predicted segmentation maps of the three best models in our evaluation metrics. The FSG-Net shows the best results, especially in segmenting thin vessels. Fig.~\ref{fig:Stage_Comprison} shows prediction results of FSG-Net obtained through deep supervision from the decoder stages indicated in Fig.~\ref{fig:neU-Net_overview}. As shown in Fig.~\ref{fig:Stage_Comprison}, intermediate predictions may not directly provide fine-grained details, but they can improve the final segmentation results due to their accurate semantic-level information and the capability to capture global context. The quantitative performance benefits of such deep supervision is further discussed in the ablation study.

In the inference settings, we padded the original image with a multiple of 32 to preserve the flexible operation of specific models. Resizing the shape can lead to informational loss, which is critical in retinal vessel segmentation that requires high-fidelity maps. When measuring the metrics, we again removed the padding to generate a perfectly similar shape to the original image with no informational loss. With this unpadding trick, metrics that require true negatives and false negatives can be decreased compared with padded or resized images. Remarkably, the majority of models examined in our study exhibited superior performance compared to their original implementations in our training settings. For example, AG-Net achieved better performance on the DRIVE dataset (cIoU:69.71, Sen:82.16) in our environment than the results reported in the original paper (cIoU:69.65, Sen:81.00). Furthermore, U-Net, despite being an early model introduced over a decade ago, demonstrated robustness by achieving middle-range performance using only pure convolution layers.

\begin{table*}[htp]
    \renewcommand{\arraystretch}{1.2}
    \centering
    \tiny
    \caption{\label{tab:cross-domain}  
        Cross-domain evaluation across datasets, DRIVE (D), CHASE\_DB1 (C), STARE (S), HRF (H). \(D \rightarrow C\) indicates that the model was trained on the (D) dataset and evaluated on the (C) dataset.
    }
    % \large
    \resizebox{0.9\textwidth}{!}{%
    \begin{tabular}{ l | c c c c c c}
        \Xhline{2\arrayrulewidth}
        \makecell{\textbf{Dataset}} & \makecell{\textbf{mIoU}} & \makecell{\textbf{F1 score}} & \makecell{\textbf{Acc}} & \makecell{\textbf{AUC}} &  \makecell{\textbf{Sen}} & \makecell{\textbf{MCC}} \\
        \hline
        \hline
        \(D \rightarrow C\) & 79.76 \textcolor{red}{(-2.92)} & 77.31 \textcolor{red}{(-3.70)} & 96.46 \textcolor{red}{(-1.05)} & 99.01 \textcolor{red}{(-0.36)} & 87.84 \textcolor{blue}{(+1.85)} & 76.14 \textcolor{red}{(-3.75)}  \\
        \(D \rightarrow S\) & 82.14 \textcolor{red}{(-3.97)} & 80.29 \textcolor{red}{(-4.81)} & 96.95 \textcolor{red}{(-0.79)} & 97.77 \textcolor{red}{(-1.19)} & 82.93 \textcolor{red}{(-3.67)} & 78.90 \textcolor{red}{(-5.05)}  \\
        \(D \rightarrow H\) & 79.24 \textcolor{red}{(-3.84)} & 76.45 \textcolor{red}{(-5.11)} & 96.21 \textcolor{red}{(-0.89)} & 98.92 \textcolor{blue}{(+0.18)} & 81.93 \textcolor{red}{(-1.68)} & 74.75 \textcolor{red}{(-5.37)}  \\
        \hline
        
        \(C \rightarrow D\) & 78.38 \textcolor{red}{(-5.68)} & 75.56 \textcolor{red}{(-7.66)} & 96.25 \textcolor{red}{(-0.79)} & 88.56 \textcolor{red}{(-9.67)} & 66.98 \textcolor{red}{(-17.2)} & 74.44 \textcolor{red}{(-7.29)}  \\
        \(C \rightarrow S\) & 81.51 \textcolor{red}{(-4.60)} & 79.54 \textcolor{red}{(-5.56)} & 96.98 \textcolor{red}{(-0.76)} & 96.34 \textcolor{red}{(-2.62)} & 77.52 \textcolor{red}{(-9.08)} & 78.09 \textcolor{red}{(-5.86)}  \\
        \(C \rightarrow H\) & 80.25 \textcolor{red}{(-2.83)} & 77.77 \textcolor{red}{(-3.79)} & 96.57 \textcolor{red}{(-0.53)} & 97.96 \textcolor{red}{(-0.78)} & 79.86 \textcolor{red}{(-3.75)} & 76.13 \textcolor{red}{(-3.99)}  \\
        \hline
        
        \(S \rightarrow D\) & 80.70 \textcolor{red}{(-3.36)} & 78.82 \textcolor{red}{(-4.40)} & 96.52 \textcolor{red}{(-0.52)} & 94.21 \textcolor{red}{(-4.02)} & 74.84 \textcolor{red}{(-9.36)} & 77.22 \textcolor{red}{(-4.51)}  \\
        \(S \rightarrow C\) & 79.87 \textcolor{red}{(-2.81)} & 77.50 \textcolor{red}{(-3.51)} & 96.56 \textcolor{red}{(-0.95)} & 99.20 \textcolor{red}{(-0.17)} & 86.04 \textcolor{blue}{(+0.05)} & 76.16 \textcolor{red}{(-3.73)}  \\
        \(S \rightarrow H\) & 78.38 \textcolor{red}{(-4.70)} & 76.01 \textcolor{red}{(-5.55)} & 96.08 \textcolor{red}{(-1.02)} & 98.79 \textcolor{blue}{(+0.05)} & 82.49 \textcolor{red}{(-1.12)} & 74.33 \textcolor{red}{(-5.79)}  \\
        \hline
        
        \(H \rightarrow D\) & 79.13 \textcolor{red}{(-4.93)} & 76.64 \textcolor{red}{(-6.58)} & 96.27 \textcolor{red}{(-0.77)} & 92.27 \textcolor{red}{(-5.96)} & 70.79 \textcolor{red}{(-13.4)} & 75.07 \textcolor{red}{(-6.66)}  \\
        \(H \rightarrow C\) & 81.80 \textcolor{red}{(-0.88)} & 79.88 \textcolor{red}{(-1.13)} & 97.10 \textcolor{red}{(-0.41)} & 99.59 \textcolor{blue}{(+0.22)} & 83.69 \textcolor{red}{(-2.30)} & 78.46 \textcolor{red}{(-1.43)}  \\
        \(H \rightarrow S\) & 82.18 \textcolor{red}{(-3.93)} & 80.15 \textcolor{red}{(-4.95)} & 97.15 \textcolor{red}{(-0.59)} & 95.06 \textcolor{red}{(-3.90)} & 76.81 \textcolor{red}{(-9.79)} & 78.89 \textcolor{red}{(-5.06)}  \\
        \Xhline{2\arrayrulewidth}
    \end{tabular}
    }
\end{table*}

To investigate the domain generalization capability of our model in retinal vessel segmentation, we conducted cross-domain validation experiments as presented in Table \ref{tab:cross-domain}. Specifically, we trained the model on the DRIVE dataset and evaluated its performance on different unseen datasets, namely CHASE\_DB1 (D $\rightarrow$ C), STARE (D $\rightarrow$ S), and HRF (D $\rightarrow$ H). As expected, performance decreased across most evaluation metrics when the model was tested on datasets different from the training domain. For example, in the D→C scenario, we observed reductions in mIoU (-2.92), F1 score (-3.70), and MCC (-3.75), although sensitivity showed a slight improvement (+1.85). Similar observations were made in other domain pairs, notably D→S and D→H, reflecting the inherent challenges posed by domain shifts such as differences in imaging modalities, resolutions, and contrast distributions.

However, despite this performance drop, the model still maintained reasonable predictive capability, indicating its robustness and capacity to learn transferable, domain-invariant features. These findings underscore the importance of enhancing domain generalization by exploring additional techniques, such as domain augmentation, pre-training strategies, linear probing, or knowledge distillation, to further improve the model's adaptability to unseen domains.

\subsection*{Ablation study}
\begin{table*}[htp]
    \renewcommand{\arraystretch}{1.2}
    \centering
    \tiny
    \caption{\label{tab:ablation-capacity}
    Ablation study on model capacity.
    }
    % \large
    \resizebox{0.9\textwidth}{!}{%
    \begin{tabular}{ l c c c | c c c}
        \Xhline{2\arrayrulewidth}
        \makecell{\textbf{model}} & \makecell{\textbf{Base\_c}} & \makecell{\textbf{Depth}} & \makecell{\textbf{Params(M)}} &  \makecell{\textbf{DRIVE}} & \makecell{\textbf{STARE}} & \makecell{\textbf{CHASE\_DB1}} \\
        \hline
        \hline
        FSG-Net-L (proposed) & 64 & [3, 3, 9, 3] & 18.32 & 83.229 & 85.100 & 81.019 \\ 
        FSG-Net-B & 64 & [2, 2, 6, 2] & 14.46 & 83.191 & 84.934 & 79.658 \\ 
        FSG-Net-S & 48 & [3, 3, 9, 3] & 10.33 & 83.145 & 84.917 & 80.529 \\ 
        FSG-Net-T & 32 & [3, 3, 9, 3] & 4.61 & 83.098 & 84.698 & 79.982 \\ 
        FSG-Net-N & 16 & [3, 3, 9, 3] & 1.17 & 82.904 & 84.692 & 79.431 \\ 
        \Xhline{2\arrayrulewidth}
    \end{tabular}
    }
\end{table*}

To further understand the impact of the model capacity and structure on FSG-Net, we conducted ablation studies. In Table \ref{tab:ablation-capacity}, we vary the depth of the down-convolution, base channel (Base\_c) and structure. The F1 score of three datasets is used as a metric here. The results showed that even the FSG-Net-N surpassed the other models. By comparing the scores in Table \ref{tab:Comparison_Metrics}, the FSG-Net-N with parameter size of 1.17M outperformed recent studies with an average rank (5.2) across all metrics on the three datasets, compared to SAU-Net's rank (10.8) with a parameter size of 0.5M, DCASU-Net's~\cite{https://doi.org/10.48550/arxiv.2202.00972} rank (8.0) with a parameter size of 2.6M, ConvU-NeXt's~\cite{HAN2022109512} rank (6.8) with a parameter size of 3.5M, and FR-UNet's rank (8.2) with a parameter size of 7.4M.

\begin{table*}[htp]
    \renewcommand{\arraystretch}{1.2}
    \centering
    \tiny
    \caption{\label{tab:ablation-module}  
    Ablation study on the proposed modules. The abbreviations represent: DC – replacement of the residual block with a down-convolution structure; GRM – integration of the GRM block; SA – spatial attention at the bottleneck; DKS – dynamic kernel size evaluation; DS – use of deep supervision.
    }
    % \large
    \resizebox{0.8\textwidth}{!}{
    \begin{tabular}{ c c c c c | c | c c c }
        \Xhline{2\arrayrulewidth}
        \textbf{DC} & \textbf{GRM} & \textbf{SA} & \textbf{DKS} & \textbf{DS} & \textbf{Params(M)} & \textbf{DRIVE} & \textbf{STARE} & \textbf{CHASE\_DB1} \\
        \hline
        \hline
        \ding{55} & \ding{55} & \ding{55} & \ding{51} & \ding{51} & 39.09 & 82.580 & 81.786 & 78.944 \\
        \ding{51} & \ding{55} & \ding{55} & \ding{51} & \ding{51} & 16.85 & 82.254 & 83.323 & 79.409 \\
        \ding{51} & \ding{51} & \ding{55} & \ding{51} & \ding{51} & 18.31 & 83.115 & 84.671 & 79.333 \\
        \ding{51} & \ding{51} & \ding{51} & \ding{51} & \ding{55} & 18.32 & 83.131 & 84.698 & 79.889 \\
        \ding{51} & \ding{51} & \ding{51} & \ding{55} & \ding{51} & 35.77 & 83.006 & 84.200 & 80.044 \\
        \ding{51} & \ding{51} & \ding{51} & \ding{51} & \ding{51} & 18.32 & 83.229 & 85.100 & 81.019 \\
        \Xhline{2\arrayrulewidth}
    \end{tabular}
    }
\end{table*}

To validate the contribution of each proposed component, we conducted a series of ablation experiments on the DRIVE, STARE, and CHASE\_DB1 datasets. Table~\ref{tab:ablation-module} presents the ablation study conducted on the main modules of FSG-Net, with each module denoted by its corresponding abbreviation. Starting from the R2U-Net~\cite{https://doi.org/10.48550/arxiv.1802.06955} baseline, chosen for its strong generalization across medical image segmentation tasks, we incrementally integrated the modules proposed in FSG-Net. As shown in the second row of Table~\ref{tab:ablation-module}, replacing the residual blocks in R2U-Net with our proposed down-convolution (as illustrated in Fig.~\ref{fig:up_down_conv}(c)) led to a slight performance drop on the DRIVE dataset, while improving the results on STARE and CHASE\_DB1. However, this modification reduced the number of parameters by more than half, indicating a favorable trade-off between efficiency and accuracy. As can be seen in the third row of the Table~\ref{tab:ablation-module}, substituting the standard U-Net-style feature concatenation with the proposed GRM block resulted in notable performance improvements on both DRIVE and STARE. This confirms that our GRM module enhances multi-scale feature aggregation and representation learning. Introducing a lightweight spatial attention mechanism at the bottleneck stage yielded consistent performance gains with only a marginal increase in parameter count. This demonstrates the module’s effectiveness in enhancing contextual understanding without significant overhead. To assess the impact of kernel size, we compared a series of three 3×3 convolutions with a single 7×7 convolution. While the 7×7 configuration exhibited minor improvements in high-resolution scenarios due to its direct abstraction style, it generally led to increased parameter counts and lower performance on other datasets, suggesting limited generalizability. Incorporating deep supervision, as shown in the sixth row of the Table~\ref{tab:ablation-module}, consistently improved performance, supporting its role in guiding the learning process through additional gradient signals during training.

\section*{Discussion}

% \begin{figure*}[ht]
%     \centering
%     \includegraphics[width=0.95\linewidth]{feature_merge2.png}
%     \caption{\label{fig:feature_merge}
%           Compared structure of feature merging in (a) FSG-Net, and (b) AG-Net.
%           }
% \end{figure*}

We previously emphasized the method of feature merging within the network architecture. This is because feature merging allows information from all stages to be mixed and provided as input to the guided filtering module. This approach can be compared to AG-Net, which also gathers features and provides them to the guided filter. Fig.~\ref{fig:feature_merge}(a) illustrates the feature merging in the proposed method, while Fig~\ref{fig:feature_merge}(b) shows the corresponding part of AG-Net. Let us focus on the left path to the guided filtering modules in both cases. As can be seen in Fig~\ref{fig:feature_merge}(a), feature maps from three stages (current, upward, and downward) are concatenated and passed through the proposed convolution block. For AG-Net, only the features from the current and lower stages are merged for the attention-guided filter. In summary, our GRM gets feature information from all stages with a modernized convolution, while the AG-Net gets it from current and lower stages with conventional convolutions. For this reason, we were able to achieve superior performance with one fewer stage than AG-Net.

The next discussion point concerns the experimental datasets. DRIVE, STARE, and Chase\_DB1 are representative datasets for retinal vessel segmentation. Models that improve the architecture of layers and attention mechanisms continue to emerge, but the number of images in such datasets is limited. Although this paper focuses on retinal vessel segmentation, it is necessary to apply the proposed model to large databases from other fields to verify its variability and scalability. To this end, we plan to extend FSG-Net to a broader spectrum of medical-imaging tasks—including semi-supervised histological image segmentation~\cite{jin2022semi,jin2024inter}, skin-lesion diagnosis and segmentation~\cite{jin2021cascade}, and COVID-19 infection delineation in CT images~\cite{jin2021domain}—so as to validate its robustness under limited annotations and pronounced domain shifts.

Finally, we discuss recent approaches to retinal vessel segmentation that differ from ours in terms of network architecture and utilization of attention mechanisms. Unlike our proposed FSG-Net, which utilizes outputs inferred from all layers of a U-Net architecture, Tan et al.~\cite{tan2024deep} proposed a W-shaped RCAR-UNet structure that aggregates outputs from the preceding U-Net and passes them forward to the next U-Net module. Meanwhile, Ding et al.~\cite{ding2024rcar} incorporated a rough set-based channel-attention mechanism into the U-Net structure to reinforce the long-range dependency of retinal vessels. This differs from our approach, which combines information from all scales and passes it through a trainable guided filter. In future studies, we plan to explore integrating these complementary attention and filtering strategies with our full-scale guided network framework, potentially yielding further improvements in segmentation performance.

\section*{Conclusion}
In this study, we presented a full-scale representation guided network (FSG-Net) for retinal vessel segmentation that demonstrates competitive performance approaching current SOTA methods on the DRIVE dataset.
A modern convolutional block tailored for retinal vessel segmentation was designed, and the guided convolution block effectively leveraged full-scale information from the feature representation network.
The proposed guided convolution block is compatible with any U-Net architecture, offering scalability for similar tasks across various domains. However, our method remains purely CNN-based, which might limit its capability compared to recent transformer-integrated U-Net architectures designed for retinal vessel segmentation tasks. Therefore, future research should explore integrating transformer modules into our full-scale feature representation approach.
We encourage future research to use the FSG-Net architecture and our experimental settings to ensure reproducibility and scalability in similar tasks.

{\small
\bibliographystyle{ieee_fullname}
\bibliography{egbib}
}

\end{document}